\documentclass[superscriptaddress,twoside,twocolumn,showpacs,preprintnumbers,amssymb,prl]{revtex4}
\usepackage{graphicx}
\usepackage{dcolumn}
\usepackage{bm}
\begin{document}
\title{Scaling Laws of Stress and Strain in Brittle Fracture}
\author{Bj{\o}rn Skjetne}
 \affiliation{Department of Chemical Engineering,
              Norwegian University of Science and Technology,\\
              N-7491 Trondheim, Norway}
 \affiliation{Department of Physics,
              Norwegian University of Science and Technology,\\
              N-7491 Trondheim, Norway}
\author{Torbj{\o}rn Helle}
 \affiliation{Department of Chemical Engineering,
              Norwegian University of Science and Technology,\\
              N-7491 Trondheim, Norway}
\author{Alex Hansen}
 \affiliation{Department of Physics,
              Norwegian University of Science and Technology,\\
              N-7491 Trondheim, Norway}
\date{\today}
\begin{abstract}
A numerical realization of an elastic beam lattice
is used to obtain
scaling exponents relevant to the extent of damage within 
the controlled, catastrophic and total regimes of mode-I brittle 
fracture. The relative fraction of damage at the onset of catastrophic 
rupture approaches a fixed value in the continuum limit. This enables 
disorder in a real material to be quantified through its relationship 
with random samples generated on the computer. 
\end{abstract}
\maketitle
Besides having wide practical relevance, 
breakdown in complex media touches on a range of 
fundamental issues involving structural disorder. 
However, 
it is only within the past two decades or so 
that adequate tools have become available which mimic how 
a truly complex material breaks~\cite{smod}. 
These tools, known as lattice models, have their origin in 
statistical physics and are especially well suited to describe 
the interplay between a continually evolving non-uniform 
stress-field and a random meso-structure.

Deviation from a perfect structure will usually 
affect the way a material breaks under strain. This is 
especially so where fibrous, porous or granular media are
concerned. Moreover, in comparisons between components made
from the same non-perfect material, sample-to-sample variations
obtained in the response of stress to strain are typically 
large when the disorder is high. In this paper we calculate
the average values of stress versus strain for a very large
number of samples to obtain the breakdown characteristics 
of materials with a given type, or strength, of disorder. 

Since the breaking characteristics of brittle 
materials depend crucially on the disorder, 
such information can be very useful in quality control. 
The size of the 
stable regime of fracture, for instance, within which an 
increasing amount of strain may be applied before catastrophic 
rupture sets in, increases with the disorder. Indeed, 
structural disorder on the microscopic level is 
often a desired trait in many materials. Consequently, 
if a relationship can be established between disorder and the 
breaking characteristics, preliminary conclusions may be drawn
about average strength and its expected variability based on
knowledge about the disorder in the material.
This is helpful in cases where experiments
are either costly or difficult to set up. With the disorder
known, ultrasound or other non-destructive probing techniques can 
be used to determine how close a
given component under stress is to complete failure. Vice versa,
with a properly calibrated lattice model detailed experimental 
knowledge of the average stress-strain response should help to
quantify the disorder and thereby predict, through simulations,
other disorder-dependent properties which might not be easy to
access experimentally. In alloys and composite materials such 
information can be used as an aid to optimize the mixture used,
for instance, with respect to a desired strength specification.

Earlier results indicate that the scaling with 
system size of the maximum force obtained in 
quasi-static brittle fracture is universal with respect to 
the disorder. This has been reported in calculations
with both the random fuse model~\cite{darc} and the 
beam model~\cite{roux,herr}. Those results, however, were obtained
at a time when numerical resources were far more limited
than today. Our calculations present evidence which is
contrary to this. Specifically, in large-scale numerical 
simulations involving a great many samples and spanning a wide 
range of disorders, the exponents which characterize scaling 
in the controlled, catastrophic and total regimes of mode-I
brittle fracture are all found to be non-universal. 
Whereas for weak disorder the exponents are strongly 
dependent on the disorder, there is a slow convergence on 
the screened percolation value in the limit of infinite 
disorder.

A universal exponent, on the other hand, 
would be very useful for predicting
strength properties since the scaling behaviour of
stress and strain would then be the same for all 
materials, regardless of the details of the disorder.
\begin{figure} [t]
\includegraphics[scale=0.68]{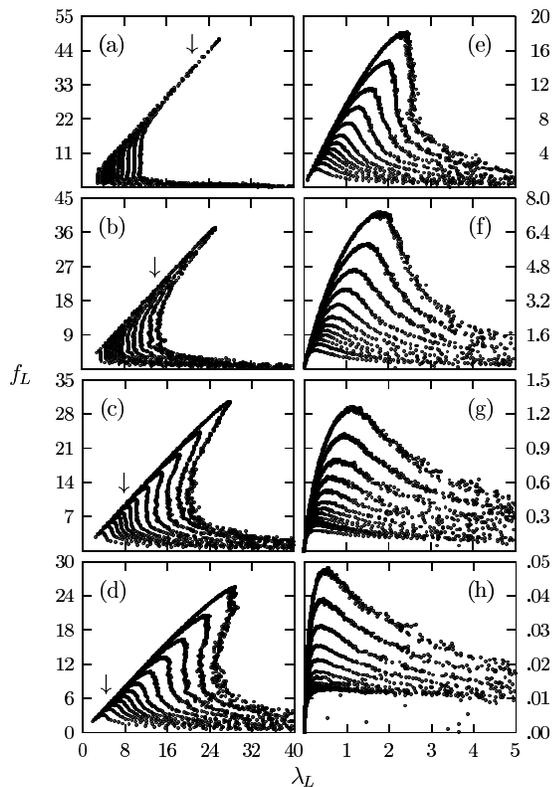}
\caption{Force $f_{L}$ versus displacement $\lambda_{L}$ obtained
         for system sizes $L=10$, 14, 17, 20, 23, 27, 32, 40, 50,
         63, 80 and~100, for a range of disorders with $D>0$. 
         On the left are shown (a)~$D=0.1$, (b)~$D=0.17$, (c)~$D=0.25$
         and (d)~$D=0.33$. On the right are shown (e)~$D=0.5$, 
         (f)~$D=1$, (g)~$D=2$ and (h)~$D=4$. Whereas in (a)--(d) the
         scale on $\lambda_{L}$ is $\times1.33f_{L}$, it varies on
         the right as (e)~$\times2.5f_{L}$, (f)~$\times5f_{L}$, 
         (g)~$\times13f_{L}$ and (h)~$\times100f_{L}$. The arrow in
         (a)--(d) shows the onset of damage for the $L=100$ system.
         \label{dplus}}
\end{figure}
An additional relationship is required 
for the exponents to be useful in the case of non-universal
scaling behaviour. Specifically, it should be possible
for the disorder in a real material to be quantified in 
terms of the same parameters as those that are used to generate
random samples on the computer.
Presently we show that, in the continuum limit, 
the fraction of damage which occurs prior to catastrophic 
breakdown approaches a fixed value which is specific 
to the disorder.

It is indeed the ease with which disorder can be included
which makes stochastic lattice models practical. The model
used in our calculations is the elastic beam model used by
Herrmann et al.\ in Ref.~\cite{herr} and Skjetne et al.\
in Ref.~\cite{skje}. The disorder is imposed on the breaking
thresholds of the beams, and the elastic properties 
are assumed to be identical from beam to beam. 
Two random sets are
generated for each lattice sample, corresponding to the
maximum breaking strengths in pure axial loading and 
pure flexure. 
The thresholds conform to the distribution which
results when a random number $r$, chosen uniformly on
the interval $0\leq r\leq 1$, is raised to a power $D$. 
This power law distribution is the most practical way to
generate multifractality, i.e., the type of behaviour which
is fundamentally associated with the existence of
scaling laws~\cite{dxdx}. 
Hence, for $D>0$ the cumulative distribution reads
\begin{equation}
    P(t)=t^{1/D},
    \label{dover0}
\end{equation}
where $0\leq t\leq 1$ is
distributed with a tail towards weak beams. For $D<0$, 
we have
\begin{equation}
    P(t)=1-t^{1/D},
\end{equation}
now with $1\leq t<\infty$ being distributed with a tail
towards strong beams. The important thing is to 
include zero or infinity within the range of 
thresholds, otherwise the distribution is 
asymptotically equivalent to no
disorder\cite{dxdx}. Both fundamental types of disorder 
are included in the present calculations. Fracture may then
be fully explored as a function of disorder simply by
varying the magnitude of $D$, with small or large
values of $|D|$ corresponding to weak or strong disorders,
respectively.

\begin{figure} [b]
\includegraphics[scale=0.54]{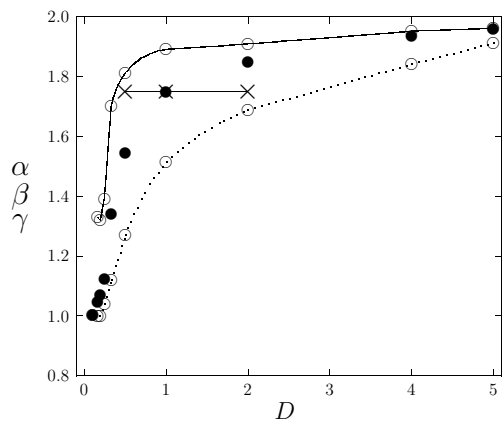}
\caption{The exponents $\alpha$ and $\beta$ for the scaling 
         with system size, $L$, of damage in the controlled (solid
         lines) and catastrophic (dotted lines) regimes, respectively,
         as a function of the disorder $D$.
         Also shown is the exponent $\gamma$ for the scaling 
         of total damage ($\bullet$) with $L$. 
         For comparison, the data from Ref.~\cite{herr}
         ($\times$), relevant to the total damage for $D=0.5$,
         $D=1$ and $D=2$, have also been included.
         \label{dgt0}}
\end{figure}
The lattice is broken by applying a uniform displacement
to the nodes defining the top row, and the first beam to
break is that which has the lowest axial strength. After
this, the breaking sequence depends on how local
stresses interact with the quenched disorder. Each time
a beam is removed from the lattice, the new mechanical 
equilibrium is obtained by minimizing the elastic energy.
In practice this is done via relaxation, i.e., using
conjugate gradients~\cite{conj}. The process by which
mechanical equilibrium is attained is assumed to be much
more rapid than the breaking of the beams, hence the
simulation emulates quasi-static fracture.

Results obtained for $D>0$ are shown in Fig.~\ref{dplus}, 
where the data points represent the average values of 
stress and strain calculated for each broken beam. The 
exponents $\alpha$, $\beta$ and $\gamma$, 
obtained for the scaling of broken beams in 
the controlled (stable), catastrophic (unstable) and 
total regimes of fracture, 
i.e., $N_{\rm S}$, $N_{\rm U}$ and $N_{\rm T}$, respectively,
have been obtained from
\begin{equation}
    N_{\rm S}\sim L^{\alpha},\hspace{2mm}
    N_{\rm U}\sim L^{\beta},\hspace{2mm}
    N_{\rm T}\sim L^{\gamma},
\end{equation}
and are shown in Fig.~\ref{dgt0}.

For weak disorders most of the thresholds are to be found in the
vicinity of the upper bound. On the application of external
force fracture therefore proceeds in a controlled manner 
only for the first few breaks. During this phase, any small
crack caused by the removal of a weak beam is prevented 
from further opening up by the stronger thresholds in the
immediate neighbourhood of the beam just broken. 
Since, at this stage, fracture is dominated by
quenched disorder rather than stress, the external 
force must now be increased
to break the next beam. A new crack will then
most likely appear away from the neighbourhood of the 
previously broken beam, especially in the case of a large
lattice. For small lattices the 
statistics of extremes dictates that the limited number of
beams present should reduce the probability of weak 
thresholds occurring. Hence, fracture is unstable from
the onset.

This cross-over in disorder between, on the one hand,
systems for which there is always a regime of stable 
crack growth (regardless of system size)
and, on the other hand, systems for which crack growth is 
unstable, or conditionally stable, from the very beginning, 
can, in fact, be identified. Using general arguments 
Roux et al.~\cite{eplx} showed that, for a fracture criterion
of the type relevant to the fuse model,
such a cross-over can be expected to occur at 
$D\sim0.5$. Hence,
for lower values of $D$ one should not expect scaling
laws to exist.

The exact value of the disorder beyond which a controlled 
regime appears
might be slightly different in the beam model, due to
a different breaking formula. Nonetheless, Fig.~\ref{dgt0} 
shows that a transition indeed occurs in the region 
$0.5\lesssim D\lesssim 1$. Although the dependence of the
exponents upon the disorder
is less pronounced beyond $D=1$, especially in the case 
of $\alpha$, the scaling is clearly non-universal, 
with the $D>1$ exponents slowly approaching the 
value $2$ in 
the limit of infinite disorder. The exponent which governs
the scaling behaviour of the total damage, however, is very 
nearly $\alpha\approx1.9$ within a wide range of disorders.

\begin{figure} [b]
\includegraphics[scale=0.54]{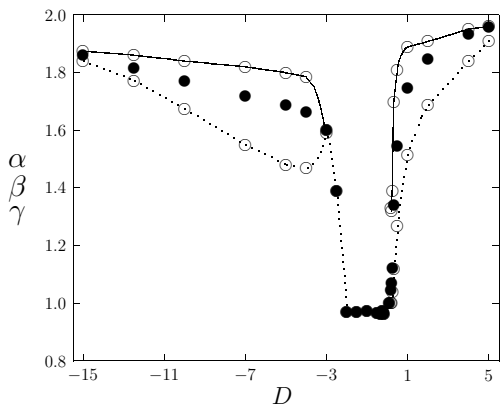}
\caption{Same as Fig.~\ref{dgt0}, but also including a
         wide range of $D<0$ disorders, where the thresholds
         are distributed with a tail towards infinitely
         strong beams. 
         \label{dal0}}
\end{figure}
In Fig.~\ref{dplus} the onset of damage for the largest 
system included, $L=100$, is shown by the small
arrow in the cases of (a)~$D=0.1$, (b)~$D=0.17$, (c)~$D=0.25$
and (d)~$D=0.33$. In (a) there is a very small controlled
regime before catastrophic rupture sets in. Stress and
strain then backtracks along the original straight-line
response towards the origin before encountering a section
where decreasing values of force correspond to the same 
displacement. The situation here is one of conditional 
stability, where a small perturbation is sufficient to 
propagate the crack further. 

\begin{figure}
\includegraphics[scale=0.68]{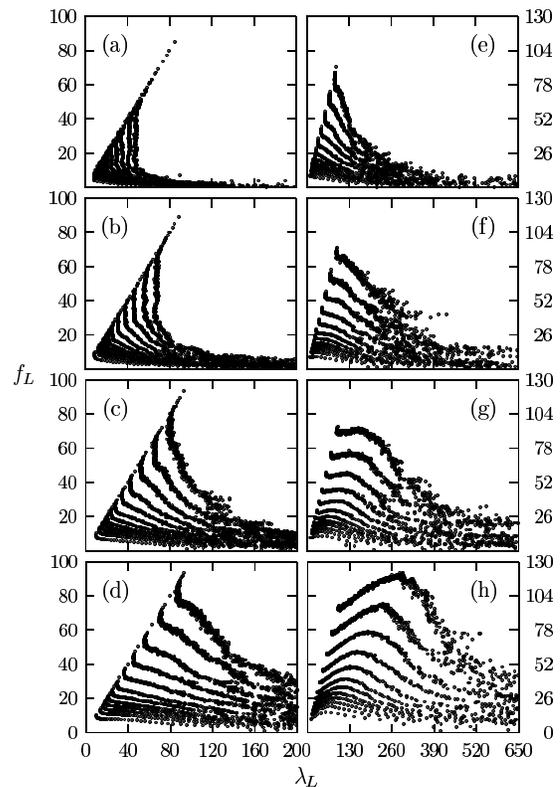}
\caption{The same as Fig.~\ref{dplus}, but for $D<0$. The plots
         on the left show (a)~$D=-0.5$, (b)~$D=-1$, (c)~$D=-1.5$
         and (d)~$D=-2$. Plots on the right show (e)~$D=-2$, 
         (f)~$D=-2.5$, (g)~$D=-3$ and (h)~$D=-4$. Note that there
         is a scale-shift in going from~(d) to~(e), the scale
         on the horizontal axes being otherwise the same. 
         \label{dminu}}
\end{figure}

Exponents relevant to fracture in the $D<0$ regime are shown 
in Fig.~\ref{dal0}, where stress-strain curves for disorders
up to $|D|=15$ have been analyzed. Stress-strain curves
between $D=-0.5$ and $D=-4$ are shown in Fig.~\ref{dminu}.
Here the initial response, a straight line from
the origin to the position of the first broken beam, is not shown. 
Fig.~\ref{dal0} shows that for small $D$ there exists a 
regime of finite extent, i.e., $-2\leq D\leq 0$, 
within which $\alpha=1$. The scaling behaviour here is trivial, 
being simply proportional to system size, and 
the first beam to break triggers catastrophic rupture. 
Within this regime the tail towards strong beams 
is not very pronounced and consequently the next 
beam in the path of the crack will most likely
be comparable in strength with that just broken.
\begin{figure} [t]
\includegraphics[scale=0.54]{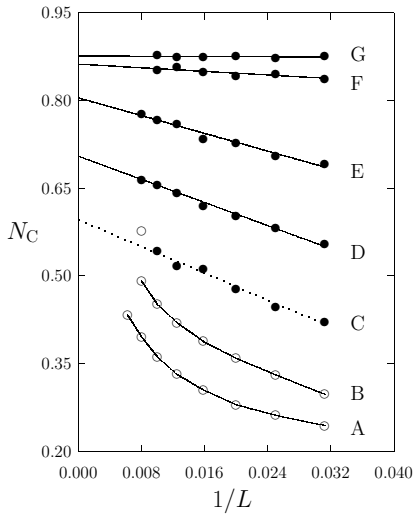}
\caption{The number of beams belonging to the controlled regime
         of fracture, $N_{\rm C}$, shown as a function of 
         inverse system size, $L^{-1}$, for (A)~$D=0.25$,
         (B)~$D=0.333$, (C)~$D=0.5$, (D)~$D=1$, (E)~$D=2$,
         (F)~$D=4$, and (G)~$D=5$. Straight 
         lines have been fit to the data in (D)--(G), and, based
         on the most significant data ($\bullet$), a
         straight line fit has also been conjectured in (C). 
         \label{dxinf}}
\end{figure}
Since the loading on 
neighbouring beams is rendered higher by its removal, crack 
growth is now localized. The case of $D=-0.5$, for instance,
is shown in plot (a) of Fig.~\ref{dminu} where fracture 
is completely unstable. Stress and strain is seen to backtrack
along the original linear response, as in the case of 
$D>0$, before encountering a region of conditional stability.
Assuming displacement control, fracture proceeds 
in a manner of slowly decreasing force until the system is
completely broken. 
\begin{figure} [b]
\includegraphics[scale=0.54]{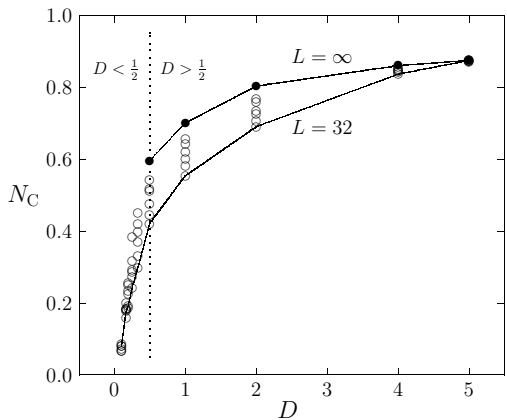}
\caption{The extent of damage at catastrophic rupture, $N_{\rm C}$,
         obtained for system sizes between $L=32$ and $L=100$,
         and extrapolated to the continuum limit, $L=\infty$, 
         based on the straight line fits in Fig.~\ref{dxinf}.
         \label{NcD}}
\end{figure}

For $D<-3$ a region of stable crack growth appears, as can be 
seen from plot (h) of Fig.\ref{dminu}. The tail towards strong
beams has now become a dominating feature in the threshold
distribution, with only a few beams remaining in the vicinity 
of the lower bound. In the early stages of fracture, therefore, 
a small crack will most likely be arrested by strong beams in 
its immediate neighbourhood. The situation is then similar to 
the case of $D>0$ in the sense that force must now be increased 
to further propagate damage.

For large $|D|$ the essential features are similar to those for 
$D>0$, with the scaling being governed by non-universal exponents. 
As before, the scaling of total damage shows the weakest
variation, with $\alpha\approx1.9$ within a very large region 
of disorders.

At the onset of catastrophic rupture the relative amount of 
damage is $\widehat{C}=N_{\rm C}/N_{\rm T}$, where $N_{\rm C}$ 
is the number of beams broken in the controlled phase and 
$N_{\rm T}$ is the total number of beams broken. In Fig.~\ref{dxinf}, 
this quantity is shown as a function of inverse system size
for $D>0$. For the sizes included in the present calculations, 
a straight-line relationship is obtained for disorders larger 
than $D=1$. With the $L=125$ data having been left out due to
low statistical significance, a line has also been fit to the 
data for $D=0.5$, although the relationship is probably not 
linear here. For disorders below $D=0.5$ the relationship is 
clearly non-linear, as can be expected based on the arguments 
of Ref.~\cite{eplx}.

The same data are shown in Fig.~\ref{NcD}, now 
however as functions
of the disorder. The result extrapolated for the continuum limit, 
corresponding to $L=\infty$, is the intersection of the
straight-line fit in Fig.~\ref{dxinf} with the vertical axis
$1/L=0$. This establishes a link between the disorder,
as generated on the computer (presently using a power-law 
distribution), and the stress-strain characteristic obtained 
in the breaking of real materials. Such a relationship is 
valuable where comparisons are sought between simulation and 
experiment. As such, the present quasi-static result for
brittle materials is mainly relevant to breakdown associated 
with fatigue, i.e., slow fracture processes.

\newpage

\end{document}